\documentclass[twocolumn,showpacs,aps,floatfix,superscriptaddress]{revtex4}
\usepackage{amsmath,amssymb,graphicx,eucal,bm}
\begin{document}
\title{Kinetics of Ring Formation}
\author{E.~Ben-Naim}
\affiliation{Theoretical Division and Center for Nonlinear
Studies, Los Alamos National Laboratory, Los Alamos, New Mexico
87545}
\author{P.~L.~Krapivsky}
\affiliation{Department of Physics,
Boston University, Boston, Massachusetts 02215}
\begin{abstract}
We study reversible polymerization of rings. In this stochastic
process, two monomers bond and as a consequence, two disjoint rings
may merge into a compound ring, or, a single ring may split into two
fragment rings. This aggregation-fragmentation process exhibits a
percolation transition with a finite-ring phase in which all rings
have microscopic length and a giant-ring phase where macroscopic rings
account for a finite fraction of the entire mass.  Interestingly,
while the total mass of the giant rings is a deterministic quantity,
their total number and their sizes are stochastic quantities.  The
size distribution of the macroscopic rings is universal, although the
span of this distribution increases with time.  Moreover, the average
number of giant rings scales logarithmically with system size.  We
introduce a card-shuffling algorithm for efficient simulation of the
ring formation process, and present numerical verification of the
theoretical predictions.
\end{abstract}
\maketitle 

\section{Introduction}

Percolation \cite{sa,gg} controls many natural processes from polymer
gelation \cite{pjf,whs,pf} and diffusion in porous media \cite{bh,ms},
to the spread of forest fires \cite{btw,bd} or infectious diseases
\cite{pg,mej,tz}. In the standard percolation picture, a system
evolves from a state with many small, microscopic, clusters into a
state with a {\em single}, macroscopic, system-spanning, cluster. This
phase transition is continuous, and it is controlled by the total
number of connections between elementary units in the system.

In this study, we show that restricting the structure of the clusters
leads to a different percolation behavior where multiple macroscopic
clusters coexist.  Percolation with multiple giant clusters has been
recently reported in theoretical studies \cite{bk05a,wd}, and it is
relevant to the production of colloidal micro-gels \cite{ms1,sv}.

Our starting point is the classic polymer gelation process introduced
by Flory \cite{pjf,whs,pf,sr,er}, a simplified model that is
essentially the mean-field theory for percolation
\cite{sa,zhe,fl,aal,jlr,bb}. In this polymerization process, a very
large number of molecular units (``monomers'') join irreversibly to
form clusters (``polymers''). This process has a second order phase
transition between a ``sol'' phase in which all polymers are finite to
a ``gel'' phase in which a {\em single} gel containing a finite
fraction of the monomers in the system emerges. With time, this gel
grows and eventually, it engulfs the entire system.

\begin{figure}[t]
\includegraphics[width=0.37\textwidth]{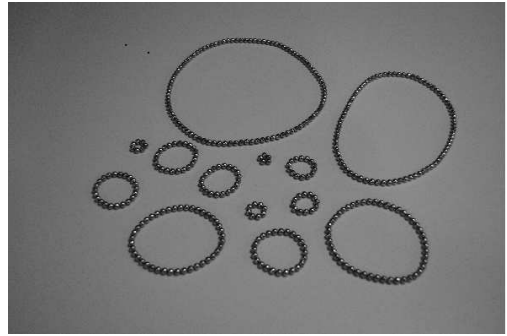}
\caption{Rings made of magnetic beads.}
\label{fig-magnetic}
\end{figure}

In the Flory model, there is no limit on the number of bonds per
monomer, and the resulting polymers may have very different
structures.  We modify the polymerization process so that all polymers
have the same structure. In our version, all monomers have exactly two
bonds, so that all polymers are rings.  Rings occur in magnetized
powders or beads \cite{kssaob} because due to magnetic interactions,
linear chains are unstable with respect to formation of rings (Figure
\ref{fig-magnetic}).  As is the case for magnetic beads, we consider 
directed rings where the bonds have directionality (Figure 
\ref{fig-aggfrag}). The results extend to undirected rings.

\section{Aggregation-Fragmentation Process}

\begin{figure}[t]
\includegraphics[width=0.35\textwidth]{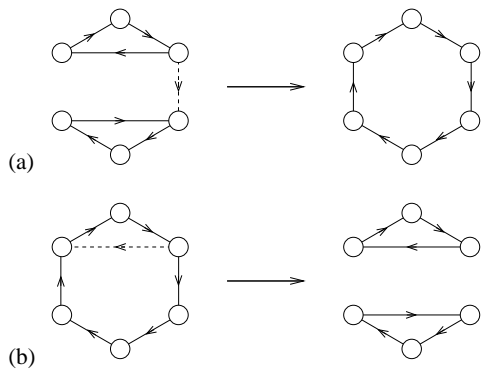}
\caption{(a) Inter-ring bonds lead to {\em aggregation}. (b)
  Intra-ring bonds result in {\em fragmentation.}}
\label{fig-aggfrag}
\end{figure}

At time $t=0$, our system consists of $N$ isolated monomers. These
particles bond to form polymeric rings through the following process.
In each elementary step, two monomers are selected at random and a
first bond is drawn between them. Subsequently, both monomers drop an
existing bond and then, the two ``dangling'' monomers form a second
bond, as shown in Figure \ref{fig-aggfrag}. Time is updated, $t\to 
t+\Delta t$ with $\Delta t=2/N$, after each step so that each monomer
experiences one bonding event per unit time.  We note that the
directionality of the first bond dictates the directionality of the
second bond. This polymerization process conserves the total number of
bonds because two bonds are gained and two bonds are lost in each
event. We assign an imaginary self-bond to every original monomer, so
that formally, the original monomers have a ring structure. Therefore,
the total number of bonds in the system equals $N$. With this
formulation, the polymerization process maintains a ring topology as
every monomer has exactly two bonds.

The above polymerization process is equivalent to an
aggregation-fragmentation process. When a monomer that belongs to ring
of size $i$ bonds with a monomer that belongs to a different ring of
size $j$, a composite ring with size $i+j$ forms (Figure
\ref{fig-aggfrag}a). Hence, rings undergo the aggregation process
\begin{equation}
\label{agg}
i,j \buildrel K_{ij} \over\longrightarrow i+j\qquad{\rm with}\qquad K_{ij}=ij.
\end{equation}
The aggregation rate $K_{ij}$ is proportional to the product of the
sizes because there are $i\times j$ distinct pairs that can bond.  We
note that the aggregation process \eqref{agg} alone constitutes the
Flory model.

A bond between two monomers in the same ring breaks that ring
into two smaller rings.  Schematically, the fragmentation process is
(Figure \ref{fig-aggfrag}b),
\begin{equation}
\label{frag}
i+j \buildrel F_{ij} \over\longrightarrow i,j \qquad{\rm with}\qquad
F_{ij}=\frac{i+j}{N}.
\end{equation}
Due to the circular symmetry, the fragmentation rate $F_{ij}$ is
proportional to the ring size, while the factor $1/N$ reflects that
for fragmentation to occur, one must pick two monomers within the same
ring. Also, the aggregation-fragmentation process specified by
\eqref{agg} and \eqref{frag} is reversible because for every
aggregation event, there is an opposite fragmentation event, and vice
verse.

Let $r_k(t)$ be the density of rings made of $k$ monomers at time
$t$. That is, if $R_k$ is the expected number of rings of size $k$,
then $r_k\equiv R_k/N$.  This size density obeys the rate equation 
\begin{eqnarray}
\label{rk-eq}
\frac{dr_k}{dt}=\frac{1}{2}\!\sum_{i+j=k}\!\!ij\,r_ir_j \!-\!kr_k
+\frac{1}{N}\left[\sum_{j>k} jr_{j}\!-\!\tfrac{k(k-1)}{2}r_k\!\right]\!.
\end{eqnarray}
The first two terms represent gain and loss due to the aggregation
process \eqref{agg}, and the last two terms represent gain and loss
due to the fragmentation process \eqref{frag}. In particular, let us
consider the two loss terms.  The total aggregation rate is, by
definition, the ring size $k$, but the total fragmentation rate,
$F_k\equiv \sum_{i+j=k}F_{ij}$, grows quadratically with size,
$F_k=\tfrac{1}{N}\binom{k}{2}=\tfrac{k(k-1)}{2N}$. Our goal is to
understand the time evolution of the density $r_k(t)$, starting from
the monomer-only initial condition, \hbox{$r_k(0)=\delta_{k,0}$}.

\section{Finite Rings}

Our implicit assumption is that the system is very large.  When
\hbox{$N\to\infty$}, one can use perturbation theory with the inverse
system size being the small parameter \cite{bk04}.  We expand the size
distribution to first order, \hbox{$r_k=c_k+\frac{1}{N}g_k+\cdots$},
and substitute this form into \eqref{rk-eq} to obtain the rate
equation
\begin{equation} 
\label{ck-eq}
\frac{dc_k}{dt}=\frac{1}{2}\sum_{i+j=k}ij c_ic_j -kc_k.
\end{equation} 
The initial condition is \hbox{$c_k(0)=\delta_{k,1}$}.  The two terms
in this equation describe gain and loss due to aggregation.  To zeroth
order, the fragmentation process is negligible because the likelihood
of picking two monomers within the same ring vanishes when
\hbox{$N\to\infty$}. Equations \eqref{ck-eq} describe the evolution of
the size distribution in the Flory model, where there is no
fragmentation. The solution to these equations is well-known 
(see \cite{fl,krb} for a review)
\begin{equation}
\label{ck}
c_k(t)=\frac{1}{k\cdot k!}(kt)^{k-1}e^{-kt}.
\end{equation} 

Let $M_n(t)=\sum_{k\geq 1}k^n c_k(t)$ be the $n$th moment of the
distribution $c_k$. The second moment diverges at a finite time as
$M_2(t)=(1-t)^{-1}$ for $t<1$, a signature of the percolation
transition at time $t=1$.  The first moment, $M_1(t)$, provides
additional information about this phase transition. Consider the
``missing mass'' \hbox{$g(t)=1-M_1(t)$}.  This quantity obeys the
transcendental equation
\begin{equation}
\label{gt}
g=1-e^{-gt}.
\end{equation}
When $t<1$, there is only the trivial solution $g=0$, and hence,
finite clusters contain all of the mass.  However, when $t>1$, there
is a second, nontrivial solution, \hbox{$0<g<1$}, and this solution is
the physical one.  Finite rings account for only a finite fraction,
$M_1=1-g$, of the total mass.  Therefore, giant rings must account for
the remaining fraction of the total mass, $g$. At time $t>1$, the
total mass of the giant rings equals $g(t)N$.

\begin{figure}[h]
\includegraphics[width=0.425\textwidth]{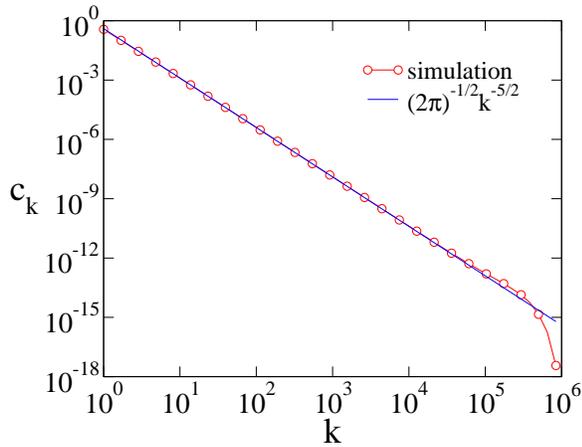}
\caption{The critical size distribution, $c_k\equiv c_k(t=1)$ versus
  $k$.  The simulation results are from $10^4$ independent
  realizations of a system with $N=10^8$.}
\label{fig-ck}
\end{figure}

At time $t=1$, the critical size distribution has a power-law tail
(Figure \ref{fig-ck}),
\begin{equation}
c_k(1)\simeq \frac{1}{\sqrt{2\pi}}k^{-5/2},
\end{equation}
when $k\gg 1$. At the critical point, the size of largest ring scales 
as $N^{2/3}$ with system size $N$ \cite{jklp,bbckw,bk05b}.

\section{Giant Rings}

When $t>1$, we expect that macroscopic rings, that is, rings that
contain a finite fraction of the total mass in the system, account for
the missing mass.  For a giant ring with size $k\propto N$, the total
aggregation rate, $k$, and the total fragmentation rate,
$\frac{k(k-1)}{2N}$, are both proportional to $N$. Hence, aggregation
and fragmentation occur with comparable rates. Also, since both rates
are proportional to the system size, aggregation and fragmentation are
very rapid. To find the size distribution of the giant rings, we
must consider the aggregation-fragmentation process governing the
giant rings.

We characterize a giant ring using the normalized size, $\ell$,
defined as $\ell=k/N$.  This quantity equals the fraction of the total
mass contained in the ring.  Let $G(\ell,t)$ be the average number of
rings with normalized size $\ell$ at time $t$. Conservation of mass
dictates
\begin{equation}
\label{cons}
g(t)=\int d\ell \,\ell\,G(\ell,t),
\end{equation}
where $g(t)$ is the nontrivial solution of \eqref{gt}.

The quantity $G(\ell,t)$ satisfies
\begin{eqnarray}
\label{gl-eq}
\frac{1}{N}\frac{\partial G(\ell,t)}{\partial t}&=&
\frac{1}{2}\int_0^\ell ds\,s (\ell-s)G(s,t)G(\ell-s,t)\nonumber\\
&-&\ell(g-\ell)G(\ell,t)\nonumber\\
&+&\int_\ell^g ds\, s\, G(s,t) -\frac{1}{2}\ell^2G(\ell,t).
\end{eqnarray}
This rate equation, essentially the continuous analog of
Eq.~\eqref{rk-eq}, describes the aggregation-fragmentation process that
governs the giant rings.  To obtain \eqref{gl-eq} from \eqref{rk-eq},
we first make the transformations $G_k\equiv Nr_k$ and $k=\ell N$, and
then, note that the aggregation loss rate is reduced by the factor
$(g-\ell)$ because self-interactions do not lead to aggregation.

\begin{figure}[ht]
\includegraphics[width=0.45\textwidth]{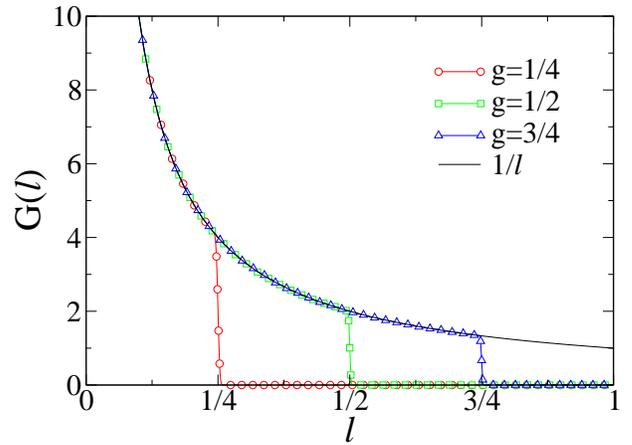}
\caption{Simulation results for $G(\ell)\equiv G(\ell,t)$ at three
  different times, $t(g=1/4)=1.150729$, $t(g=1/2)=1.386294$, and
  $t(g=3/4)=1.848392$. The time $t(g)=\frac{1}{g}\ln \frac{1}{1-g}$
  follows from \eqref{gt}.  Also shown for reference is the
  theoretical prediction \eqref{gl}. The simulations results are from
  $10^7$ independent realizations of a system with size $N=10^6$.}
\label{fig-gl}
\end{figure}

Since the total aggregation and fragmentation rates are both
proportional to the system size, the left-hand side of \eqref{gl-eq}
is negligible in the large-$N$ limit. We therefore replace the
left-hand-side of Eq.~\eqref{gl-eq} with zero to determine the {\em
time-dependent} distribution $G(\ell,t)$. The resulting non-linear
integral equation has the remarkably simple solution (see Figure
\ref{fig-gl})
\begin{equation}
\label{gl}
G(\ell,t)=
\begin{cases}
\ell^{-1}&\ell<g(t),\\
0       & \ell>g(t).
\end{cases}
\end{equation}
Indeed, this solution obeys the mass conservation statement
\eqref{cons}. Surprisingly, the size distribution is universal,
although the span of the distribution grows with time,
\hbox{$0<\ell<g(t)$}. Therefore, at time $t>1$, there are giant rings
of all sizes up to the maximal value $g(t)N$.

The distribution of rings includes two distinct components: $Nc_k$
gives the average number of finite rings, and $G(\ell)$ gives the
average number of giant rings.  Of course, the former expression
applies at all times, while the latter holds only for $t>1$.  The
giant rings grow at the expense of the finite rings and eventually,
they take over the entire system as $g\to 1$ when $t\to\infty$.

Finite rings and giant rings undergo separate, essentially decoupled,
aggregation-fragmentation processes. Indeed, the rate equation
\eqref{ck-eq} for $c_k$ is closed, while the rate equation
\eqref{gl-eq} for $G(\ell)$ is, in practice, also a closed
equation. There is a constant flux of mass, $N\times dg/dt$, from
finite rings to giant rings, and this flux couples the two types of
rings. This coupling enters only through the fraction $g(t)$ which
appears explicitly in \eqref{gl-eq}.  

\begin{figure}[t]
\includegraphics[width=0.45\textwidth]{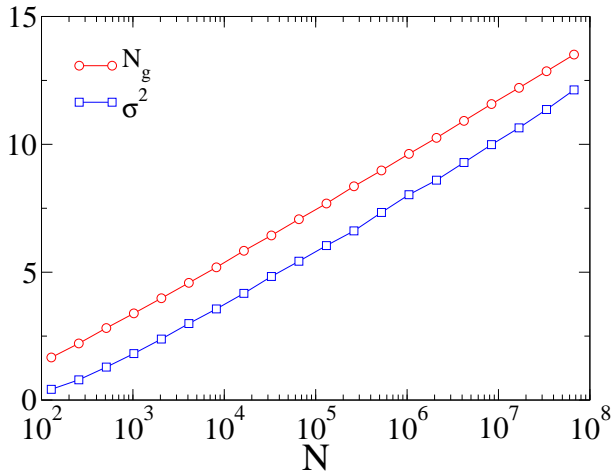}
\caption{The average number of giant rings, $N_g$, and the variance
  $\sigma^2$, versus system size $N$. The simulation results represent
  an average over $10^5$ independent realizations. We measured $N_g$
  and $\sigma^2$ by counting the number of rings with size $k>4\ln N$
  at time $t=2$.}
\label{fig-ng}
\end{figure}

The distribution \eqref{gl} implies that there are multiple giant
rings: the average number of giant rings, $N_g$, scales
logarithmically with system size (Figure \ref{fig-ng})
\begin{equation}
\label{Ng}
N_g\simeq \ln N.
\end{equation}
This behavior follows from $N_g=\int_{l_*}^g d\ell\, G(\ell)$. The
lower limit $\ell_*=k_*/N$ can be deduced from the criterion
$N\sum_{k\geq k_*}c_k(t)=1$ that estimates the size of the largest
{\em finite} ring.  Using this criterion together with Eq.~\eqref{ck}
we find $k_*\simeq (t-\ln t -1)^{-1}\ln N$ and therefore \hbox{$\ell_*\sim
N^{-1}\ln N$}.

Since the merger-breakup process is random, we expect that the variance in
the number of giant rings, $\sigma^2$, is proportional to the mean,
$\sigma^2\simeq \ln N$. Numerical simulations confirm this behavior
(Figure \ref{fig-ng}). Hence, the standard deviation
\begin{equation}
\label{sigma}
\sigma\simeq \sqrt{\ln N}
\end{equation}
quantifies  fluctuations in the number of giant rings.

\begin{figure}[t]
\includegraphics[width=0.425\textwidth]{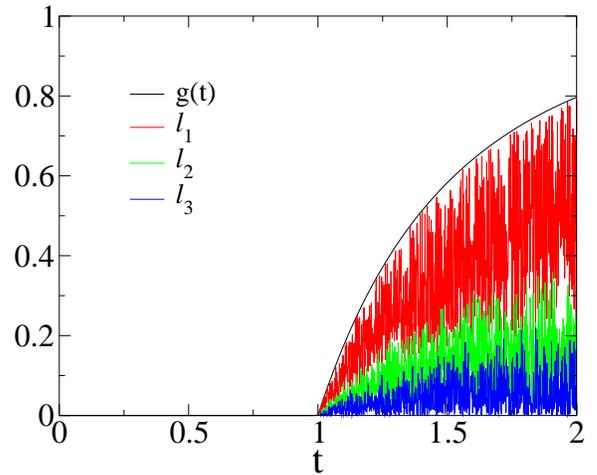}
\caption{The largest three rings. Shown is the time evolution of
  $\ell_n(t)$, the size of the $n$th largest ring at time $t$, for
  $n=1,\ 2,\ 3$.  The results are from a single run of a system with
  $N=10^6$.  Also shown is the cumulative mass $g(t)$.}
\label{fig-lt}
\end{figure}

Figure \ref{fig-lt} shows the normalized sizes of the three largest
rings as a function of time using a simulated system. These sizes
exhibit huge fluctuations as giant rings constantly merge and break on
a very fast time scale.  Interestingly, while the size of an
individual giant ring is a stochastic quantity, the total size of all
giant rings, $g(t)$, is a deterministic quantity.

The number of finite rings is proportional to $N$, while the number of
giant rings scales only logarithmically with $N$. Equation \eqref{ck}
shows that monomers dominate in the long-time limit. By comparing the
average number of monomers, $Nc_1=Ne^{-t}$, with the the average
number of giant rings, given by \eqref{Ng}, we conclude that giant
rings overtake finite rings when $t\gg t_f$ with
\begin{equation}
\label{tf}
t_f\simeq \ln N.
\end{equation}
In writing this expression, we ignored secondary logarithmic
corrections.

For times $t\gg t_f$, the ring-size distribution reaches a steady
state.  Setting $g=1$ in \eqref{gl} shows that $N_k$, the average
number of rings with (unnormalized) size $k$, has the following form
\begin{equation}
\label{nk}
N_k=\frac{1}{k},
\end{equation}
for all $1\leq k\leq N$. Thus, at the steady state, there are rings of
all lengths, from finite rings to macroscopic rings.  The distribution
\eqref{nk} also follows from the detailed-balance condition
$K_{ij}c_ic_j=F_{ij}c_{i+j}$ with the solution $c_k=(Nk)^{-1}$
\cite{krb,lt}.

\section{Shuffling Algorithm}

Throughout this paper, we presented results of Monte Carlo simulations
that support the theoretical predictions. We implemented an elegant algorithm
that takes advantage of an isomorphism between the polymerization
process and a card shuffling process. In the card shuffling algorithm
\cite{swb,fow,ad,pd}, we start with an ordered deck of $N$ cards,
numbered 1 through $N$. Then, at each elementary step, we pick two
cards at random and exchange their positions.  For example, the first
two step in shuffling a deck of $6$ cards may look like
\begin{equation*}
1\underline{2}34\underline{5}6 \to 15\underline{3}\underline{4}26
\to 154326\to \cdots.
\end{equation*}
Time is updated by $\Delta t=2/N$ after each step, \hbox{$t\to
  t+2/N$}, and thus, each card participates in one shuffling event per
unit time, on average.

\begin{figure}[t]
\includegraphics[width=0.425\textwidth]{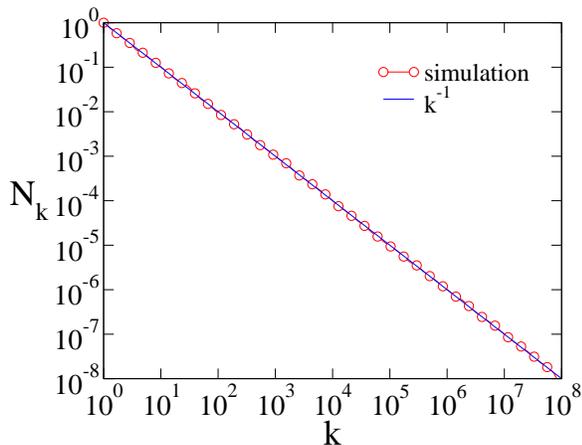}
\caption{The average number $N_k$ of rings of size $k$ at the steady-state.
  The simulation results are from $10^3$ independent
  realizations of a system with $N=10^8$ at time $t=20$.}
\label{fig-nk}
\end{figure}

We now use cycles to represent permutations. For example, six cards
ordered $134265$ are represented by $(1)(234)(56)$ because there are
three cycles: the card $1$ forms a cycle of length one, the cards
$234$ form a cycle of length three, and the cards $56$ form a cycle of
length two.  Initially, there are $N$ cycles of length $1$. Then,
exchange of two cards in distinct cycles leads to merger, while
exchange of two cards in the same cycle leads to breakup. For example,
the following steps generate the merger and breakup events in Figure
\ref{fig-aggfrag},
\begin{equation*}
(1\underline{2}3)(4\underline{5}6)\to  (156423), \quad{\rm and}\quad
(1\underline{5}64\underline{2}3)\to  (123)(456).
\end{equation*}
Furthermore, the merger rate and the breakup rate are given by
\eqref{agg} and \eqref{frag}. Hence, the dynamics of cycles in the
shuffling process are identical to the dynamics of rings in the
polymerization process.

The above algorithm is straightforward and efficient.  The shuffling
steps take ${\cal O}(N)$ operations per unit time, and moreover,
tracing the cycle structure requires only ${\cal O}(N)$
operations. This linear algorithm enabled us to simulate large systems
with as many as $N=10^8$. Figure \ref{fig-nk} demonstrates the
excellent agreement between the simulation results and the theoretical
prediction \eqref{nk}.

The distribution $N_k$ given in \eqref{nk} equals the average number
of cycles of length $k$ for a random permutation of $N$ elements
\cite{dek,mb}. As expected, repeated shuffling randomizes the card
order and according to \eqref{tf}, the number of exchanges required to
generate a perfectly random shuffle scales as $N\ln N$.

A natural generalization is to $n$-card shuffling where $n$ randomly
chosen cards are reordered according to a prescribed rule.  For
example, if $n=3$, we may follow the cyclic rule $123\to 231$.  The
equivalent polymerization process now involves merger of $n$ polymers.
Straightforward generalization of the Flory model shows that the total
gel mass, $g(t)$, satisfies \cite{jg}
\begin{equation}
\label{gtn}
1-g=e^{-\frac{1-(1-g)^{n-1}}{(n-1)!}t},
\end{equation}
with $0<g<1$ in the giant-ring phase $t>(n-2)!$. We anticipate that
the distribution of giant rings is given by \eqref{gl}, except that
the total mass is now specified by \eqref{gtn}. Our numerical
simulations of the three-card process confirm this behavior.

\medskip
\section{Discussion}

In summary, we studied a ring polymerization process in which a bond
between two monomers results either in aggregation of two rings into
one or in fragmentation of one ring into two. This process exhibits a
percolation transition with a finite-ring phase in which all rings are
microscopic and a giant-ring phase in which multiple macroscopic rings
coexist. While the cumulative mass of the giant rings is
deterministic, the sizes of individual giant rings are stochastic.
Moreover, the giant rings exhibit huge fluctuations due to the
extremely rapid merger and breakup processes.  Finally, the size
distribution of giant rings is stationary, although the span of this
distribution grows with time.

The aggregation-fragmentation process that governs the rings is
perfectly {\em reversible}. On the one hand, the distribution of ring
size reaches a stationary state where detailed balance is formally
satisfied. On the other hand, this final distribution is not
thermodynamic because the number of rings varies logarithmically,
rather than linearly, with system size. Phase transitions with
non-thermodynamic states were previously observed only in {\em
  irreversible} aggregation-fragmentation processes
\cite{kr,mkb,bk08}.

The ring formation process can be generalized in many ways. We
focused on the mean-field version, and it will be interesting to study
two-dimensional rings where spatial correlations play an important
role. Another direction for further study is percolation of polymers
with other types of fixed structures, for example, polymers where all
monomers have exactly three bonds \cite{pjf}.

Finally, we notice that the unusual behaviors in the giant-ring phase
are the consequence of the basic topological constraint, namely that
the polymers maintain a ring structure. This suggests to investigate
the influence of other constraints such as planarity \cite{cky,ce}.
Another interesting question is what happens if the polymers are
membranes \cite{membranes} such as spheres (say due to the surface
tension) that can merge and divide.

\medskip

We thank Kipton Barros for useful discussions and Talia Ben-Naim for
experimenting with magnetic rings.  This research is supported by DOE
grant DE-AC52-06NA25396 and NSF grant CCF-0829541.


\begin{thebibliography}{99}

\bibitem{sa}     D.~Stauffer and A.~Aharony,
                 {\em Introduction to Percolation Theory}
                 (Taylor \& Francis,  London, 1992).

\bibitem{gg}     G.~Grimmett, 
                 {\em Percolation} 
                 (Springer, Berlin, 1999).

\bibitem{pjf}    P.~J.~Flory,
                 J. Amer. Chem. Soc. {\bf 63}, 3083 (1941).

\bibitem{whs}    W.~H.~Stockmayer, 
                 J. Chem. Phys. {\bf 11}, 45 (1943).

\bibitem{pf}     P.~J.~Flory,
                 {\em Principles of Polymer Chemistry}
                 (Cornell University Press, Ithaca, 1953).

\bibitem{bh}     A.~Bunde and S.~Havlin (editors), {\em Percolation and
                 Disordered Systems: Theory and Applications}, Physica
                 A {\bf 366}, (1998).
            
\bibitem{ms}     M.~Sahimi,
                 {\em Flow and Transport in Porous Media and Fractured Rock}
                 (VCH, Boston, 1995).

\bibitem{btw}    P.~Bak, C.~Tang, and K.~Wiesenfeld, 
                 Phys. Rev. Lett. {\bf 59}, 381 (1987).

\bibitem{bd}     B.~Drossel and F.~Schwabel, 
		 Phys. Rev. Lett. {\bf 69}, 1629 (1992). 

\bibitem{pg}     P.~Grassberger,
                 Math. Biosci. {\bf 63}, 157 (1983). 

\bibitem{mej}    M.~E.~J.~Newman,
                 Phys. Rev. E {\bf 66}, 016128 (2002).
          
\bibitem{tz}     T.~Tome and R.~M.~Ziff,  
                 Phys. Rev. E {\bf 82}, 051921 (2010).

\bibitem{bk05a}  E.~Ben-Naim and P.~L.~Krapivsky, 
                 J. Phys. A {\bf 38}, L417 (2005); 
                 J. Phys. Cond. Matter. {\bf 17}, S4249 (2005).

\bibitem{wd}     W.~Chen and R.~M. D'Souza, 
                 arxiv:1011.5854.

\bibitem{ms1}    M.~J.~Murray and M.~J.~Snowden,
                 Adv. Coll. Int. Sci. {\bf 54}, 73 (1995).

\bibitem{sv}     B.~R.~Saunders and B.~Vincent,
                 Adv. Coll. Int. Sci. {\bf 80}, 1 (1999).

\bibitem{sr}     R.~Solomonoff and A.~Rapaport,
                 Bull. Math. Biophys. {\bf 13}, 107 (1959).

\bibitem{er}     P.~Erd\H os and A.~R\'enyi,
                 Publ.\ Math.\ Inst.\ Hungar.\ Acad.\ Sci. {\bf 5}, 17 (1960).

\bibitem{zhe}    R.~M.~Ziff, E.~M.~Hendriks, and M.~H.~Ernst, 
                 Phys. Rev. Lett. {\bf 49}, 593 (1982).

\bibitem{fl}     F.~Leyvraz, 
                 Phys. Rep. {\bf 383}, 95 (2003).

\bibitem{aal}    A.~A.~Lushnikov, 
                 Phys. Rev. Lett. {\bf 93}, 198302 (2004). 

\bibitem{jlr}    S.~Janson, T.~\L uczak and A.~Rucinski,
                 {\em Random Graphs} (John Wiley \& Sons, New York, 2000).

\bibitem{bb}     B.~Bollob\'as,
                 {\em Random Graphs} (Academic Press, London, 1985).

\bibitem{kssaob} K.~Kohlstedt, A.~Snezhko, M.~V.~Sapozhnikov, 
                 I.~S.~Aranson, J.~S.~Olafson, and E.~Ben-Naim,
                 Phys.\ Rev.\ Lett. {\bf 95}, 068001 (2005).

\bibitem{bk04}   E.~Ben-Naim and P.~L.~Krapivsky, 
                 J. Phys. A {\bf 37}, L189 (2004).

\bibitem{krb}    P.~L.~Krapivsky, S. Redner and E. Ben-Naim,
                 {\it  A Kinetic View of Statistical Physics}
                 (Cambridge University Press, Cambridge, 2010)

\bibitem{jklp}   S.~Janson, D.~E.~Knuth, T.~\L uczak, and B.~Pittel,
                 Rand. Struct. Alg. {\bf 3}, 233 (1993).

\bibitem{bbckw}  B.~Bollob\'as, C.~Borgs, J.~T.~Chayes, J.~H.~Kim, 
                 and D.~B.~Wilson, 
                 Rand.\ Struct.\ Alg. {\bf 18}, 201 (2001).

\bibitem{bk05b}  E. Ben-Naim and P. L. Krapivsky, 
                 Phys. Rev. E {\bf 71}, 026129 (2005).

\bibitem{lt}     D.~A.~Lowe and L.~Thorlacius, 
                 Phys. Rev. D {\bf 51}, 665 (1995). 

\bibitem{swb}    S.~W.~Golomb,
                 SIAM Rev. {\bf 4}, 293 (1961).
                
\bibitem{fow}    L.~Flatto, A. Odlyzko, and D.~Wales, 
                 Ann. Prob. {\bf 13}, 151 (1985). 

\bibitem{ad}     P.~Diaconis and D.~Aldous, 
                 Amer. Math. Month. {\bf 93}, 333 (1986). 

\bibitem{pd}     P.~Diaconis, 
                 Proc. Natl. Acad. Sci. USA {\bf 93}, 1659 (1996). 

\bibitem{dek}    D.~E.~Knuth, 
                {\it The Art of Computer Programming, vol. 3: Sorting and 
                Searching} (Addison-Wesley, New York, 1998).
       
\bibitem{mb}     M.~B\'ona,
                 {\it Combinatorics of Permutations} 
                 (Chapman and Hall, Boca Raton, 2004).

\bibitem{jg}     J.~Jiang and H.~Gang, 
                 Phys. Rev. B {\bf 39}, 4659 (1989).

\bibitem{kr}     P.~L.~Krapivsky and S.~Redner, 
                 Phys.\ Rev.\ E {\bf 54}, 3553 (1996).
  
\bibitem{mkb}    S.~N.~Majumdar, S.~Krishnamurthy, and M.~Barma,
                 Phys. Rev. Lett. {\bf 81}, 3691 (1998).

\bibitem{bk08}   E. Ben-Naim and P. L. Krapivsky, 
                 Phys. Rev. E {\bf 77}, 061132 (2008).

\bibitem{cky}    C.~Godr\`eche, I.~Kostov, and I.~Yekutieli, 
                 Phys. Rev. Lett. {\bf 69}, 2674 (1992).

\bibitem{ce}     P.~Collet and J.-P.~Eckmann,
                 J. Stat. Phys. {\bf 121}, 1073 (2005).

\bibitem{membranes} 
                {\it Statistical Mechanics of Membranes and Interfaces}, 
                ed. D.~R.~Nelson, T.~Piran and S.~Weinberg 
                (World Scientific, Singapore, 1989).

\end{thebibliography}
\end{document}